\documentclass[preprint,superscriptaddress,aps,longbilbiography,floatfix]{revtex4-2}
\usepackage{xcolor}
\usepackage{graphicx}
\usepackage{bm}
\usepackage{amsmath}
\usepackage{amssymb}
\usepackage{color}
\usepackage{float}
\usepackage{multirow}

\begin{document}

\title{Competing ferromagnetic and antiferromagnetic interactions in non-altermagnetic Ru\textsubscript{1-x}Cr\textsubscript{x}O\textsubscript{2}}

\author{David T. Plouff}\email{dplouff@udel.edu}
\affiliation{Department of Physics and Astronomy, University of Delaware, Newark, DE, USA}
\author{Nawsher J. Parvez}
\affiliation{Department of Physics and Astronomy, University of Delaware, Newark, DE, USA}
\author{Subhash Bhatt}
\affiliation{Department of Physics and Astronomy, University of Delaware, Newark, DE, USA}
\author{Xixiang Zhang}
\affiliation{Physical Science and Engineering Division, King Abdullah University of Science and Technology (KAUST), Thuwal, Saudi Arabia}
\author{Adam A. Aczel}
\affiliation{Neutron Scattering Division, Oak Ridge National Laboratory, Oak Ridge, TN, USA}
\author{Jonathan Gaudet}\email{jonathan.gaudet@nist.gov}
\affiliation{NIST Center for Neutron Research, National Institute of Standards and Technology, Gaithersburg, MD, USA}
\affiliation{Department of Materials Science and Eng., University of Maryland, College Park, MD, USA}
\author{John Q. Xiao}\email{jqx@udel.edu}
\affiliation{Department of Physics and Astronomy, University of Delaware, Newark, DE, USA}

\date{\today}

\begin{abstract}
We investigate the proposal of hole doping RuO\textsubscript{2} via alloying with Cr ions to induce altermagnetism. 
Thin film samples of Ru\textsubscript{1-x}Cr\textsubscript{x}O\textsubscript{2} ($0 \leq x \leq 0.28$) were prepared by reactive magnetron co-sputtering on TiO\textsubscript{2} substrates, with epitaxial nature verified by X-ray diffraction and composition depth profiles determined by X-ray photoelectron spectroscopy combined with Ar etching. 
Neutron diffraction measurements of samples with $x = 0$ and $x = 0.23$ at low temperatures show no evidence of long-range altermagnetic order with spins oriented along the $c$-axis or within the $ab$-plane. 
In contrast, temperature dependent susceptibility measurements in samples with $x \geq 0.16$ show a gradual downturn below $T\approx20~$K, suggestive of antiferromagnetic interactions, although this coexists with ferromagnetic hysteresis and remnant magnetization at $4~$K for samples with $x \geq 0.23$. 
Together, our magnetometry and neutron scattering measurements suggest the coexistence of antiferromagnetically coupled ferromagnetic CrO\textsubscript{2} clusters. This indicates that Cr ions do not hole dope Ru bands to induce an altermagnet state, but rather the holes remain localized to the Cr ions. 
\end{abstract}

\maketitle

\section{\label{sec:level1}Introduction\protect}

In the field of spintronics the most essential device is the magnetic tunnel junction (MTJ), which utilizes metallic ferromagnet (FM) electrodes for their ability to conduct spin currents and host an easily switchable magnetic state \cite{julliere1975tunneling,parkin2004giant,yuasa2004giant}. 
Although in recent years there has been a thrust towards replacing the FM components with antiferromagnet (AFM) materials, owing to their faster spin dynamics and elimination of the stray magnetic field, the general absence of spin polarized currents and the extreme difficulty in switching the AFM state (N\'eel vector) have been a significant roadblock for implementing AFM's in MTJ's \cite{baltz2018antiferromagnetic,jungwirth2016antiferromagnetic}. 
The most promising path for overcoming these challenges is to use the recently discovered class of materials known as altermagnets, which have a non-relativistic spin-split band structure that alternates in momentum space and coexists with compensated magnetic ordering, thus yielding the ideal material properties of both FM's and AFM's combined \cite{vsmejkal2022emerging,shao2024antiferromagnetic}. 
The spin splitting in altermagnets arises from an anisotropic crystal field with a rotation symmetry connecting the two spin sublattices, which breaks the $PT$ symmetry that otherwise enforces spin degenerate bands \cite{vsmejkal2020crystal,yuan2020giant,mazin2021prediction,krempasky2024altermagnetic}. 
The crystal rotation symmetry of an altermagnet determines the parity of the spin-splitting, with $d$-wave ($C_{4}$) and $g$-wave ($C_{6}$) being the most common.
In particular, a thin-film $d$-wave altermagnetic metal has the most immediate applications in spintronic devices because it has a non-zero spin conductivity tensor, in contrast to $g$-wave altermagnets in which it globally cancels to zero \cite{gonzalez2021efficient,belashchenko2025giant}.

RuO\textsubscript{2} is a metal that is easily synthesized as a thin film and crystallizes in a rutile structure that has the $C_{4}$ rotation symmetry needed to produce $d$-wave spin splitting. 
It has been the subject of thorough investigation as an altermagnet candidate material; however, the underlying antiferromagnetic order, which is a prerequisite for altermagnetism, has been questioned.
In 2017, shortly before the discovery of altermagnetism, a neutron diffraction experiment suggested that RuO\textsubscript{2} has a $c$-axis AFM state with a weak magnetic moment of $0.05~\mu_B$/Ru \cite{berlijn2017itinerant}. This was corroborated by a resonant elastic x-ray scattering (REXS) experiment which also found signatures of $c$-axis AFM \cite{zhu2019anomalous}. 
Both of these results were later shown to be artifacts of the measurement, with the apparent magnetic Bragg peak in neutron scattering resulting from a double scattering phenomena, and the REXS peak resulting from the anisotropic crystal structure itself \cite{kessler2024absence,occhialini2026structural,gregory2025resonant}. 
The absence of magnetic moment on the Ru sites has been additionally confirmed by highly sensitive muon spin rotation experiments \cite{kessler2024absence,hiraishi2024nonmagnetic}. 
Furthermore, the most direct probe for altermagnetism would be the observation of spin splitting in the band structure or the Fermi surface by angle-resolved photoelectron spectroscopy (ARPES) or quantum oscillations of resistance, which have both been performed and found to be consistent with a non-magnetic ground state \cite{liu2024absence,osumi2026spin,wu2025fermi,kim2026definitive,qian2026determining}.
These results are all in agreement with the past consensus that RuO\textsubscript{2} is a Pauli paramagnet, as determined by magnetic susceptibility measurements \cite{Guthrie_1931,Fletcher1968,ryden1970magnetic,paul2025growth,kiefer2025crystal,paul2026fermi}.
Nonetheless, after the initial reports of an RuO\textsubscript{2} AFM state, a wide range of studies searching for altermagnetic features in RuO\textsubscript{2} were performed by many indirect measurements, such as exchange bias \cite{abel2025probing,fields2026non}, various transport properties such as the anomalous Hall effect \cite{feng2022anomalous,tschirner2023saturation,jeong2025metallicity}, spin-to-charge conversion by spin-pumping \cite{wang2026absence,wang2024inverse,yang2026ferromagnetic}, spin-torque generation \cite{bai2022observation,karube2022observation,bose2022tilted,guo2024direct,zhang2025electrical,qiao2026enhanced}, and terahertz emission spectroscopy \cite{liu2023inverse,plouff2025revisiting,jechumtal2025spin,cao2025insight}. These indirect measurements provided conflicting conclusions, highlighting the need to perform direct measurements of altermagnetism rather than relying on indirect probes.

Despite stoichiometric RuO\textsubscript{2} being nonmagnetic, it remains a reasonable question whether it can be induced into an altermagnetic state. 
Density functional theory has suggested that hole doping Ru bands can induce an AFM state with $0.4$ holes per Ru site, which can be achieved via $\approx 10\%$ Ru vacancies or $\approx 20\%$ Cr substitution on Ru sites \cite{smolyanyuk2024fragility,smolyanyuk2025origin}.
Considering that CrO\textsubscript{2} is also a rutile structured metal, hosts a collinear ferromagnetic ordering, and has 2 fewer $d$ electrons than RuO\textsubscript{2}, it is a good candidate for hole doping while maintaining the proper crystal structure and metallicity.
Two previous experimental papers have studied the magnetic properties of thin film Ru\textsubscript{1-x}Cr\textsubscript{x}O\textsubscript{2} alloys, one restricted to $x \geq 0.56$ and the other restricted to $x \leq 0.3$ \cite{west2015magnetic,wang2023emergent}. 
Both papers agreed that for $x \geq 0.3$, a ferrimagnetic state was formed, in accordance with a trend towards the ferromagnetic state of pure CrO\textsubscript{2}. 
In Ref. \cite{wang2023emergent}, $x=0.2$ was found to have zero remnant magnetization coexisting with a zero-field anomalous Hall effect, and the results were interpreted as originating from altermagnetism with a N\'eel vector along the $[110]$ direction.
These results were disputed by a subsequent theoretical paper, which instead attributed the anomalous Hall response to ferromagnetic clusters of CrO\textsubscript{2}, which are antiferromagnetically coupled and indicate that moments are localized to Cr ions and do not effectively hole dope the Ru bands \cite{smolyanyuk2025origin}. 
Considering the potential impact of an RuO\textsubscript{2} based $d$-wave altermagnet, it is worthwhile to investigate the existence of a long range AFM state induced by Cr alloying using a direct measurement technique such as neutron diffraction.

We investigate epitaxial Ru\textsubscript{1-x}Cr\textsubscript{x}O\textsubscript{2} thin films with $0 \leq x \leq 0.28$ to determine whether Cr substitution stabilizes altermagnetism. 
The crystal structure was confirmed to be epitaxial by X-ray diffraction (XRD), and the composition throughout the depth was determined by X-ray photoelectron spectroscopy (XPS) combined with Ar etching. 
Temperature dependent magnetic susceptibility and low temperature hysteresis measurements on Cr-alloyed samples revealed magnetic behavior inconsistent with an altermagnetic state. 
Neutron diffraction measurements were carried out on samples with $x = 0$ and $x = 0.23$ to directly probe for long-range altermagnetism. No magnetic Bragg peaks that relate to breaking the four-fold screw symmetry were observed in either sample, indicating the absence of a long-range altermagnetic state. 
These results instead support the existence of ferromagnetic CrO\textsubscript{2} clusters that are coupled antiferromagnetically. 
These findings suggest that Cr alloying of non-magnetic RuO\textsubscript{2} is not a viable route toward realizing a thin film $d$-wave altermagnetic metal.

\section{Methods}

Thin films of Ru\textsubscript{1-x}Cr\textsubscript{x}O\textsubscript{2} with $0\leq x\leq 0.28$ were deposited on TiO\textsubscript{2} substrates by reactive magnetron co-sputtering. The  base pressure was $5\times10^{-8}~$Torr for pure RuO\textsubscript{2} films and $1\times10^{-6}~$Torr for Cr-alloyed films. For pure RuO\textsubscript{2} ($x=0$), the substrate temperature was 600 $^{\circ}$C, and for alloyed films the substrate temperature was 450 $^{\circ}$C. In both cases the substrates were baked for 10 min prior to deposition to reach equilibrium temperature. Pure RuO\textsubscript{2} films were deposited at a total pressure of 15 mTorr with an Ar/O\textsubscript{2} ratio of 5:1 using 200 W RF power applied to a pure Ru target. Alloyed films were deposited at a total pressure of 3 mTorr with the same 5:1 Ar/O\textsubscript{2} ratio. A fixed RF power of 175 W was applied to a Cr target, while the RF power applied to the Ru target was varied from 18-24 W to obtain different Cr concentrations.

The crystal structural of the films was characterized by symmetric and asymmetric XRD measurements, including $\theta/2\theta$ scans and reciprocal space maps (RSMs). Measurements were made using a Rigaku Smartlab diffractometer equipped with a Ge (220) double-bounce monochromator. 

The composition of alloyed films throughout the film depth was characterized by XPS measurements in a Thermo Fisher Scientific K-Alpha XPS Spectrometer system with ion-beam etching capability. The CasaXPS software program was used to analyze the Ru 3p\textsubscript{3/2} and Cr 2p\textsubscript{3/2} peaks, which were obtained by the average of five high resolution scans between $450-600~$eV with step size of 0.2 eV and dwell time of 30 ms. The X-ray spot size was $400~\mu$m in diameter and the ion beam spot size was 2 mm x 3 mm, using a 2 keV monatomic ion beam with an approximate etch rate of $0.12~$nm/s. Three unique spots were measured at each depth, and the composition is reported as the average value from the three spots with the error taken as the standard deviation of the three measurements. The overall film composition was calculated as the average composition throughout the film thickness.

Magnetization was measured as a function of field and temperature using a Quantum Design 12T DynaCool Physical Property Measurement System. 
The DC magnetic susceptibility ($\chi$) was measured in $1~$T magnetic field following both zero-field cooling (ZFC) and field cooling (FC). Because we found negligible differences between the two measurement protocols, the ZFC and FC susceptibilities were averaged to improve the signal-to-noise ratio. An example of the ZFC and FC scans is shown in the supplemental material (SM). 
Identical measurement sequences were performed on bare TiO\textsubscript{2} substrates, and the resulting substrate contribution was subtracted from the measured sample signal to isolate the magnetic response of the film. 
For Cr-alloyed samples, the susceptibility $\chi$ was fitted using the Curie-Weiss equation with a temperature-independent $\chi_0$ term \cite{mugiraneza2022tutorial}. From the fit we extracted the Curie-Weiss temperature ($\theta_{CW}$) and the effective moment ($\mu_{eff})$.

Neutron diffraction measurements were acquired on samples with compositions $x = 0$ and $x = 0.23$ using the thermal neutron triple-axis spectrometer VERITAS at the High Flux Isotope Reactor (HFIR) in Oak Ridge National Laboratory. The $x = 0$ samples were measured at $100~$K, while the $x = 0.23$ samples were measured at $10~$K using a closed cycle refrigerator. To increase counting statistics, multiple samples were stacked and co-aligned for each measurement. The samples were aligned with their $b$-axis oriented vertically, so perpendicular to the scattering plane. This configuration allowed us to probe diffraction within the $(h0l)$ scattering plane using $2.37~$\AA\ incident neutrons, which was confirmed by measuring the substrate nuclear Bragg peaks. Following the alignment, the film's (002) nuclear Bragg reflection was measured to further validate the orientation. It's Bragg intensity was then used to determine the normalization constant and served to estimate the magnetic Bragg scattering of a putative altermagnetic state. Finally, because they are predicted to be the brightest magnetic peaks, we scanned for magnetic scattering at the (100) and (001) positions.

\section{Results and Discussion}

\begin{figure*}
\centering \includegraphics[width=1\columnwidth]{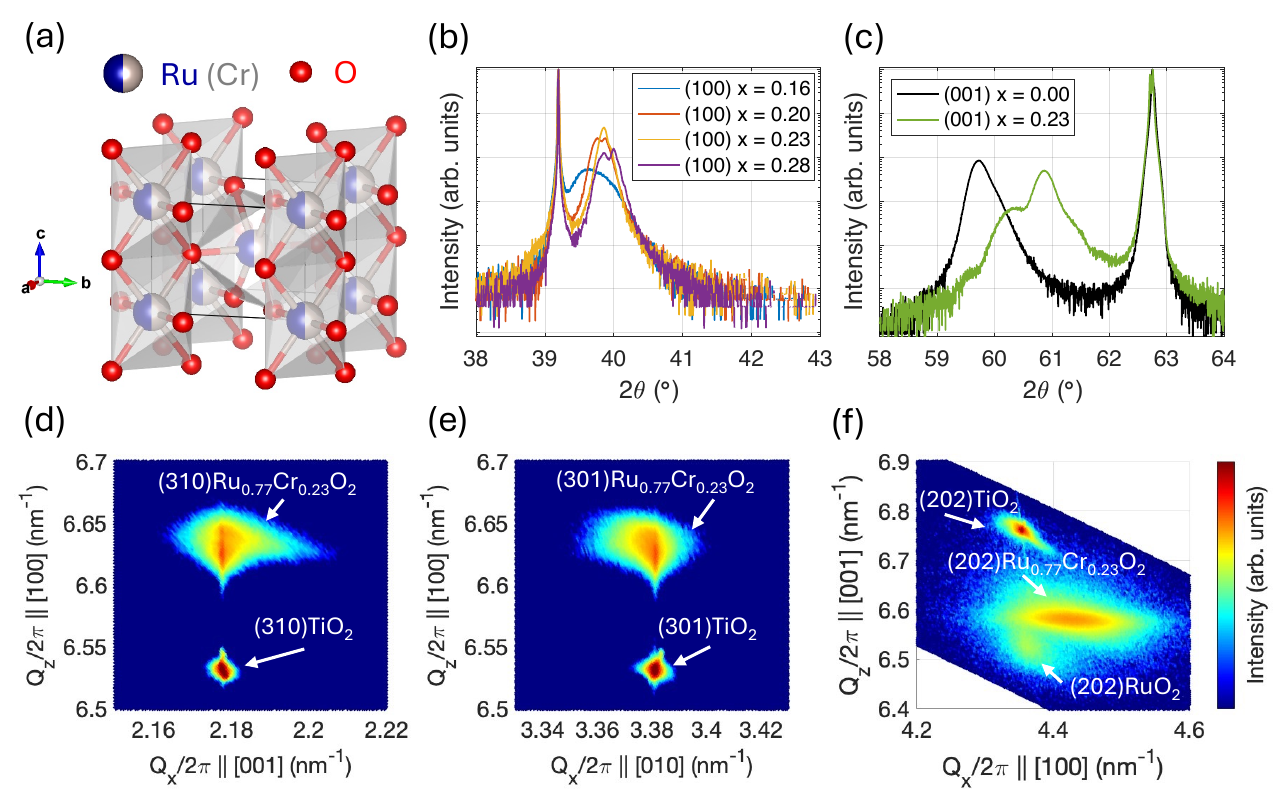}
\caption{\textbf{Crystal structure characterization:} (a) rutile unit cell with Ru or Cr atoms at the octahedral sites, (b) \& (c) $\theta/2\theta$ scans of (100) and (001) oriented samples respectively, (d) \& (e) reciprocal space maps of the (310) and (301) planes in the (100) oriented sample with $x=0.23$, and (f) reciprocal space map of the (202) planes in the (001) oriented sample with $x=0.23$.}
\label{fig:1}
\end{figure*}

\subsection{Structural and chemical characterization by X-ray diffraction and photoelectron spectroscopy}

We first confirm the crystal quality of the films using XRD, which is summarized in Figure~\ref{fig:1}. Fig. \ref{fig:1}~(a) shows the rutile unit cell (image produced using the VESTA software \cite{momma2011vesta}), which is formed by both RuO\textsubscript{2} and CrO\textsubscript{2}. Ru\textsubscript{1-x}Cr\textsubscript{x}O\textsubscript{2} films of various thicknesses (see SM Table I) were grown on rutile TiO\textsubscript{2} substrates with (100) and (001) orientations. Symmetric $\theta/2\theta$ scans for both orientations are shown in Fig. \ref{fig:1}~(b) and (c),  with the legend entries indicating the film composition determined by XPS. For both orientations, the film peak shifts systematically to higher angles with increasing Cr content, indicating a reduction in the lattice parameters consistent with substitution of Ru by the smaller Cr ion. 

The $\theta/2\theta$ scans confirm that the out-of-plane orientation of the films matches that of the substrate. To probe the in-plane structure, asymmetric RSMs were measured. Fig. \ref{fig:1}~(d) and (e) show RSMs of the (310) and (301) peaks in the (100) oriented film with $x = 0.23$, while Fig. \ref{fig:1}~(f) shows the (202) peak of the (001) oriented sample with the same composition. The (100) oriented samples are highly strained, with in-plane lattice parameters clamped to the substrate (see SM for RSMs of additional samples). A $20~$nm RuO\textsubscript{2} buffer layer was introduced for the (001) oriented sample to promote partial relaxation of the alloyed layer. The relaxed nature of the (001) oriented film provides sufficient separation between film and substrate reflections, enabling resolution of the film peak in the neutron diffraction measurement.

\begin{figure}
\centering \includegraphics[width=0.5\columnwidth]{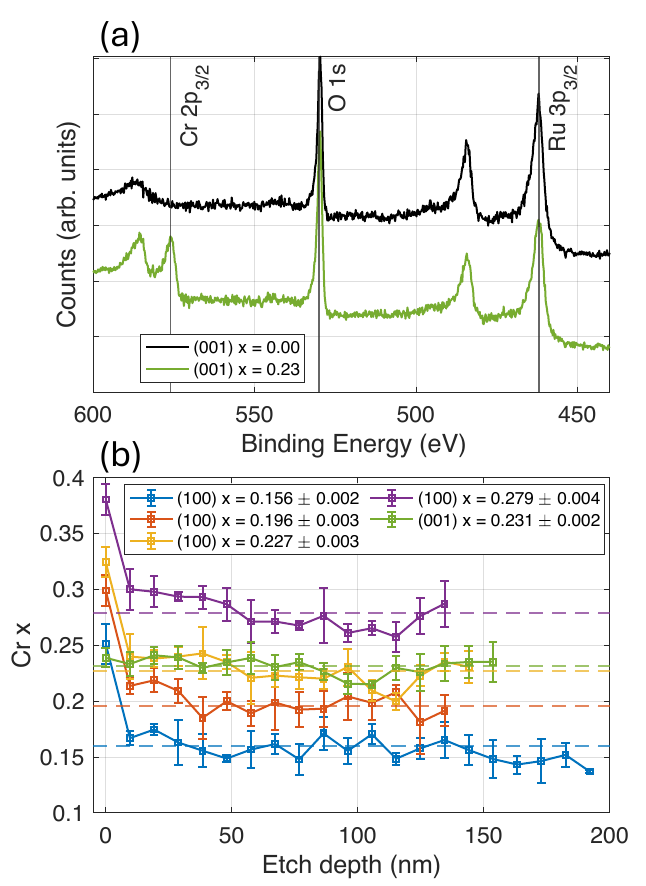}
\caption{\textbf{Elemental composition characterization:} (a) representative XPS spectra from the (001) oriented samples used for neutron scattering, with vertical shift applied for clarity, and (b) XPS depth profiles of the elemental fraction of Cr $x$ for each sample, with the color matched dashed lines indicating the samples mean value of $x$ throughout the depth.}
\label{fig:2}
\end{figure}

The chemical composition of the films was characterized by XPS depth profiles, as shown in Figure~\ref{fig:2}. Fig. \ref{fig:2}~(a) shows representative XPS spectra from the $x = 0$ and $x = 0.23$ samples after one etch cycle, while Fig. \ref{fig:2}~(b) shows the Cr concentration $x$ throughout the depth for each sample (excluding $x = 0$). For the (100) oriented samples, a Cr-rich surface layer is observed prior to etching; however, the composition becomes relatively uniform throughout the depth after the initial etch cycle. 

\subsection{Magnetic hysteresis and susceptibility}

\begin{figure}
\centering \includegraphics[width=0.5\columnwidth]{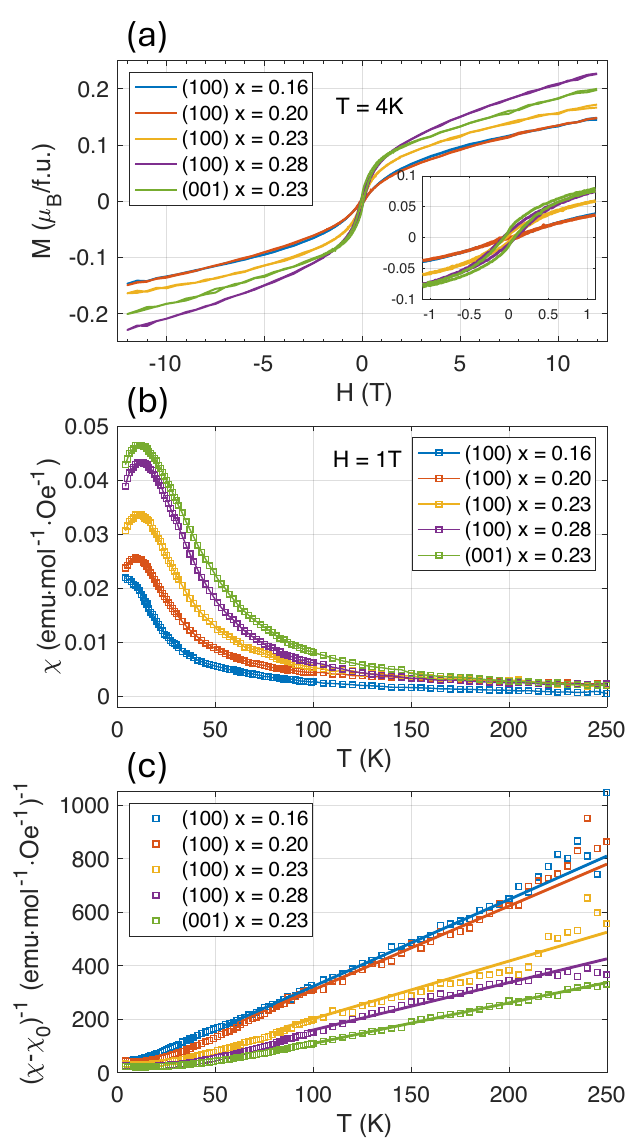}
\caption{\textbf{Magnetic hysteresis and susceptibility:} (a) $12~T$ wide magnetization vs field hysteresis loops measured at $4~K$. (b) Magnetic susceptibility per mol as a function of temperature, measured in $1~T$ magnetic field, and (c) the inverse magnetic susceptibility of each sample with a linear fit line of the same color. In all panels, the magnetic field was applied along $[001]$ for (100) oriented samples and along $[110]$ for the (001) oriented sample.}
\label{fig:3}
\end{figure}

Magnetic hysteresis and susceptibility measurements are shown in Figure~\ref{fig:3}. For all measurements, the applied magnetic field was oriented in the film plane along the $[001]$ and $[110]$ directions for the (100) and (001) oriented samples respectively. The magnetic hysteresis loops measured up to $12~$T at $4~$K are shown in Fig. \ref{fig:3}~(a) after subtraction of the substrate background. An inset of the low-field range in Fig. \ref{fig:3}~(a) shows that all samples with $x \geq 0.23$ have a small remanent magnetization and finite coercivity. None of the samples reaches magnetic saturation up to $12~$T, and the induced moment at $12~$T generally increases with Cr concentration. The (100) oriented samples with $x = 0.16$ and $x = 0.20$ exhibit nearly identical moments, while the (001) oriented sample with $x = 0.23$ has a larger induced moment than its (100) oriented counterpart. The molar magnetic susceptibility is shown in Fig. \ref{fig:3}~(b). All samples exhibit a gradual downturn in susceptibility below approximately $20~$K, although no well-defined transition temperature is observed. 

The susceptibility $\chi$ was fit using the Curie-Weiss equation plus a temperature-independent $\chi_0$ term in the high temperature regime, and the corresponding linearized inverse susceptibility $(\chi-\chi_0)^{-1}$ is shown in Fig. \ref{fig:3}~(c). The extracted Curie-Weiss temperatures $\theta_{CW}$ of the (100) samples were found to be $0.1~$K, $1.2~$K, $6.6~$K, and $10~$K, for $x = 0.16, 0.20, 0.23,$ and $0.28$ respectively, while it was $28.1~$K for the (001) sample with $x = 0.23$. Positive values of $\theta_{CW}$ indicate predominantly ferromagnetic exchange interactions, whereas values near zero indicate weak magnetic interactions. Accordingly, the (100) oriented samples with $x = 0.16$ and $x = 0.20$ exhibit only weak ferromagnetic interactions, while the larger values of $\theta_{CW}$ for higher Cr concentrations indicate increasingly strong ferromagnetic exchange despite the absence of a clear ferromagnetic ordering transition.

\begin{figure}
\centering \includegraphics[width=0.5\columnwidth]{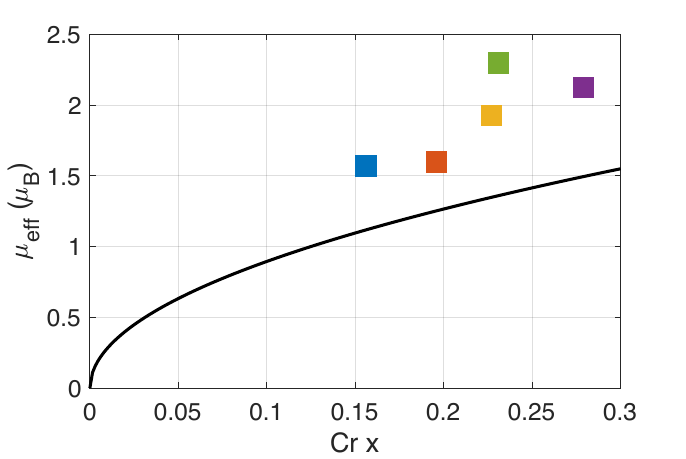}
\caption{\textbf{Composition dependence of  effective moment:} extracted effective moment from Curie-Weiss fitting, compared with effective moment from model of single Cr ions plotted in black.}
\label{fig:4}
\end{figure}

The extracted effective moments are plotted as a function of Cr concentration in Figure~\ref{fig:4}, showing a systematic increase in the effective moment with increasing $x$. Compared with the effective moment expected for diluted spin-1 Cr ions, all extracted effective moments are significantly larger than the single-ion prediction. This enhancement suggests the presence of ferromagnetically coupled Cr clusters, which produce an  effective moment exceeding that of diluted-ions. We speculate that the gradual decrease in susceptibility seen for each sample in Fig. \ref{fig:3}~(b) is related to the onset of antiferromagnetic coupling between such ferromagnetic clusters of Cr-ions, which may occur in a similar temperature range as ferromagnetic interactions.

\subsection{Neutron diffraction}

\begin{figure*}
\centering \includegraphics[width=1\columnwidth]{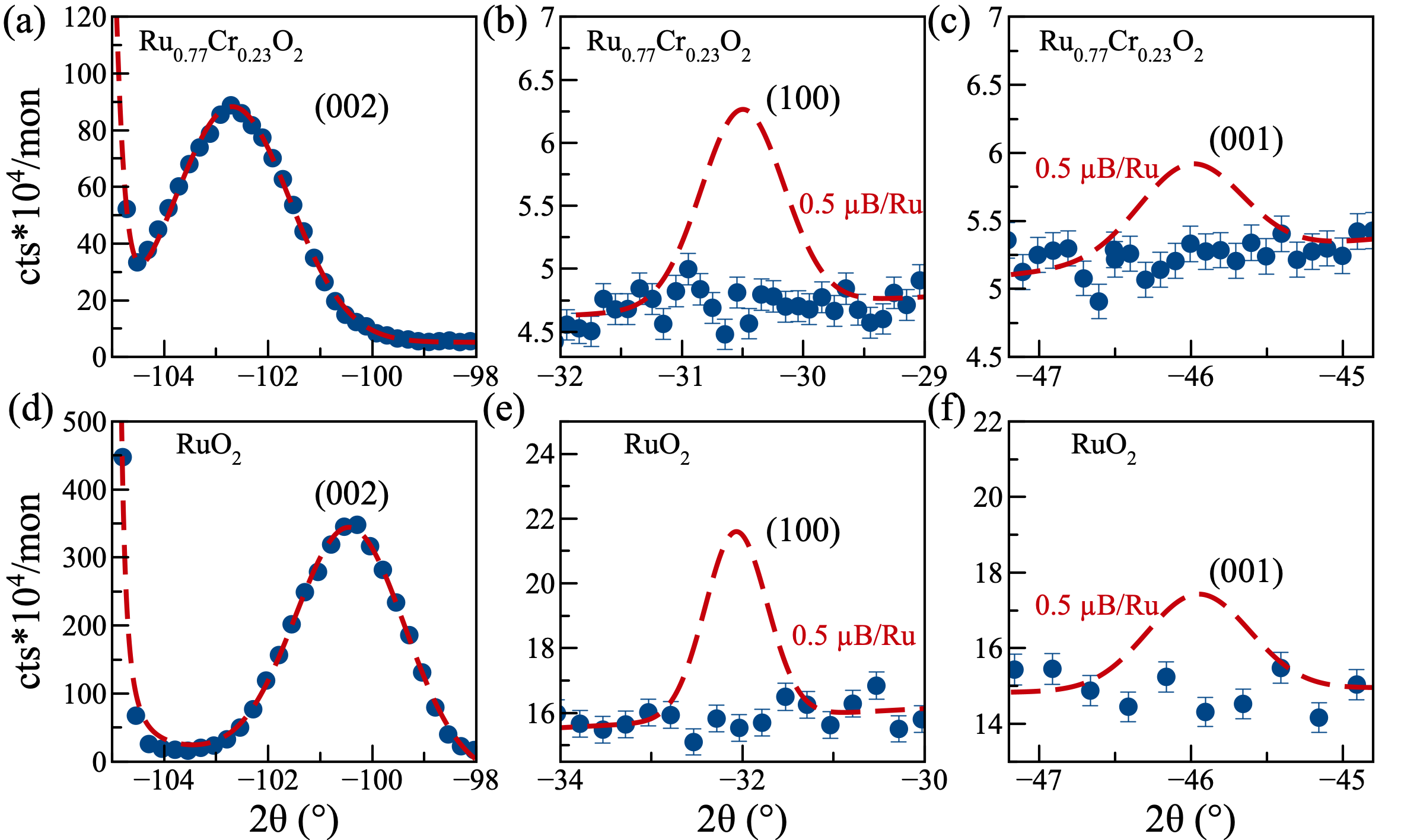}
\caption{\textbf{Neutron diffraction of Ru\textsubscript{0.77}Cr\textsubscript{0.23}O\textsubscript{2} and RuO\textsubscript{2}:} measurements of the $x = 0.23$ sample in panels (a)-(c), where panel (a) is the nuclear (002) film peak, (b) is the (100) magnetic Bragg peak location, and (c) is the (001) magnetic Bragg peak location. Measurements of the same peaks in the $x = 0$ sample are shown in panels (d)-(f) with the same measurement ordering. For all panels, the statistical error for each intensity point corresponds to 1 standard deviation. }
\label{fig:5}
\end{figure*}

Neutron diffraction of the (001) oriented films with $x = 0$ and $x = 0.23$ is reported in Figure~\ref{fig:5}. The samples were aligned in the $(h0l)$ scattering plane using the substrate reflections, after which the structurally allowed film (002) nuclear Bragg peak was measured as shown in Fig. \ref{fig:5}~(a) and (d). The film (002) reflection is well resolved from the much stronger substrate peak owing to the film's relaxed lattice parameters. Consistent with the XRD measurements, the (002) peak of the $x = 0.23$ sample is shifted to a higher scattering angle relative to the $x = 0$ sample, reflecting the reduction in lattice parameters with increasing Cr concentration. The integrated intensity of the film (002) reflection was used to normalize the calculated magnetic Bragg peak intensity.  

Magnetic scattering was then measured around the location of the structurally forbidden (100) and (001) reciprocal lattice positions, which should have strong intensity for an altermagnetic state. The $x = 0.23$ sample was measured at $10~$K while the $x = 0$ sample was measured at $100~$K. The results are shown in Fig. \ref{fig:5}~(b) and (c), and Fig. \ref{fig:5}~(e) and (f) respectively. In each panel, the dashed curves represent the calculated magnetic Bragg peak of an altermagnetic state with an ordered moment size of $0.5~\mu_B$ per formula unit. We estimated the magnetic intensity from normalization to the measured (002) nuclear reflection. The $0.5~\mu_{B}$ value was chosen because it is approximately the size of the ordered moment in Cr-alloyed RuO\textsubscript{2} predicted by recent density functional theory calculations in Ref. \cite{smolyanyuk2025origin}. No magnetic Bragg peaks are observed at either location for either composition, indicating that any ordered moment is substantially smaller than the predicted value.

These measurements provide no evidence for long-range altermagnetism in either sample composition. It should be noted that magnetic neutron scattering is sensitive only to the component of ordered moment which is orthogonal to the scattering vector. Consequently, the (100) peak is sensitive to moments along either the $c$-axis or in the $ab$-plane, whereas the (001) peak is sensitive to moments only in the $ab$-plane. The absence of magnetic scattering at both Bragg positions thus rules out long-range altermagnetic order for the expected moment orientations.

\section{Conclusion}

We have synthesized epitaxial Ru\textsubscript{1-x}Cr\textsubscript{x}O\textsubscript{2} thin films with $0 \leq x \leq 0.28$ and probed their magnetic ground state by magnetometry and neutron diffraction. Low temperature hysteresis measurements reveal ferromagnetic signatures such as remanent magnetization and finite coercivity for samples with $x \geq 0.23$, while in general all samples have an induced moment at $12~$T that increases with Cr concentration. Magnetic susceptibility measurements of all Cr-alloyed samples showed a gradual downturn below approximately $20~$K, although there is no clear transition temperature, which we presume to be related to antiferromagnetic interactions. These results are complemented by neutron diffraction measurements which show the absence of a (100) or (001) magnetic Bragg peak in $x = 0$ sample, and demonstrates for the first time that $x = 0.23$ also does not support the sought-after $d$-wave altermagnetic state. Instead, these results are consistent with antiferromagnetically coupled ferromagnetic clusters originating from Cr-rich regions with localized moments. This implies that Cr-alloying into RuO\textsubscript{2} does not effectively hole dope Ru bands, which is necessary to develop an altermagnetic state. Future efforts to realize a thin film $d$-wave altermagnetic metal should go beyond the Ru\textsubscript{1-x}Cr\textsubscript{x}O\textsubscript{2} material platform.

\section*{Author contribution}
D.T.P. and J.Q.X. conceived the project. 
D.T.P. deposited the films and characterized them by XRD and VSM. N.J.P and D.T.P. measured and analyzed the film composition by XPS. N.J.P. and S.B. helped D.T.P develop the co-sputtering recipe. D.T.P., J.G., and A.A.A. measured and analyzed the elastic neutron scattering data. All authors discussed the results and provided feedback.

\section*{Competing interests}
The authors declare no conflict of interest.

\section*{Data availability}
All data are available upon request from dplouff@udel.edu.

\section*{Acknowledgments}
This research was sponsored by NSF DMR-2316664, NSF through the University of Delaware Materials Research Science and Engineering Center (MRSEC), DMR-2011824, DOE EPSCoR DE-SC0024284, and King Abdullah University of Science and Technology (KAUST), ORFS-2022-CRG11-5031.2. 
The authors acknowledge the use of facilities at the University of Delaware, including the Materials Growth Facility (MGF) which is partially supported by the National Science Foundation Major Research Instrumentation under Grant No.1828141 and UD-CHARM, a National Science Foundation MRSEC, under Award No. DMR-2011824, and also the use of the Surface Analysis Facility (SAF) XPS which is sponsored by National Science Foundation, Major Research Instrumentation, Award Number: CHE-1428149. A portion of this research used resources at the High Flux Isotope Reactor, a DOE Office of Science User Facility operated by the Oak Ridge National laboratory. The beamtime was allocated to VERITAS on proposals number IPTS-34132 and IPTS-33591. The support for neutron scattering was provided in part by the Center for High-Resolution Neutron Scattering, a partnership between the National Institute of Standards and Technology and the National Science Foundation under Agreement No. DMR-2010792. The identification of any commercial product or trade name does not imply endorsement or recommendation by the National Institute of Standards and Technology.  


\clearpage
\newpage
\section*{Supplemental Material}

Wide-angle XRD gonio scans of each sample are reported in Figure \ref{suppfig:1}. In Fig. \ref{suppfig:1}(a) each (100) oriented sample with different $x$ is reported to show there are no additional out-of-plane crystal grains. In Fig. \ref{suppfig:1}(b) and (c) the (001) oriented samples with $x = 0$ and $x = 0.23$ are reported respectively, with labels S1, S2, etc. indicating the sample number. These panels illustrate that the multiple samples which were stacked during the neutron measurements were structurally identical, which is consistent with their simultaneous deposition.

\setcounter{figure}{0}
\renewcommand{\figurename}{FIG.}
\renewcommand{\thefigure}{S\arabic{figure}}

\begin{figure*}[h]
\centering \includegraphics[width=1\columnwidth]{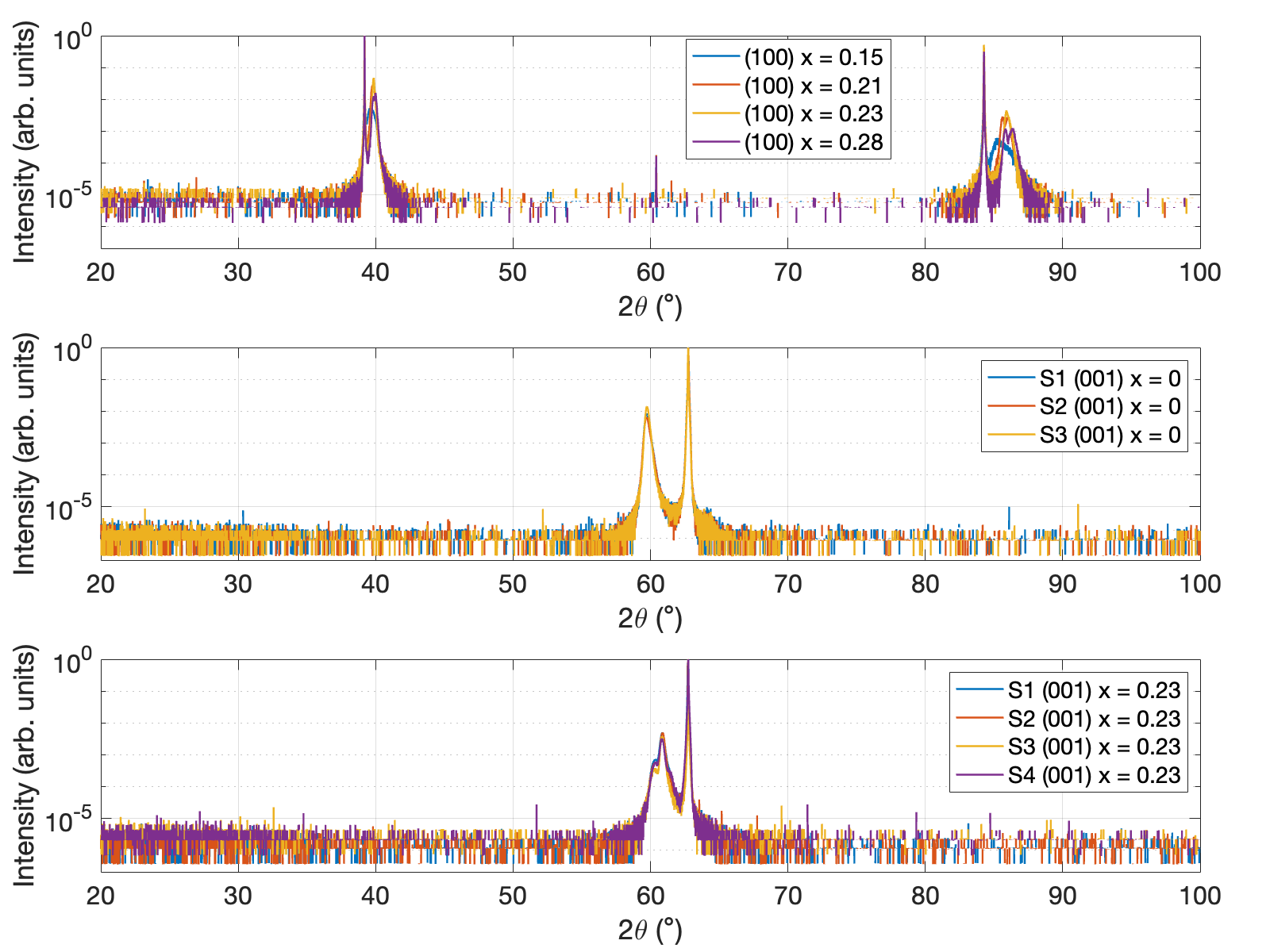}
\caption{\textbf{Wide-angle XRD $\theta/2\theta$ scans of all samples:} (a) (100) samples with each $x$ value, (b) all 3x samples with (001) $x=0$, and (c) all 4x samples with (001) $x=0.23$.}
\label{suppfig:1}
\end{figure*}

\begin{figure*}[h]
\centering \includegraphics[width=1\columnwidth]{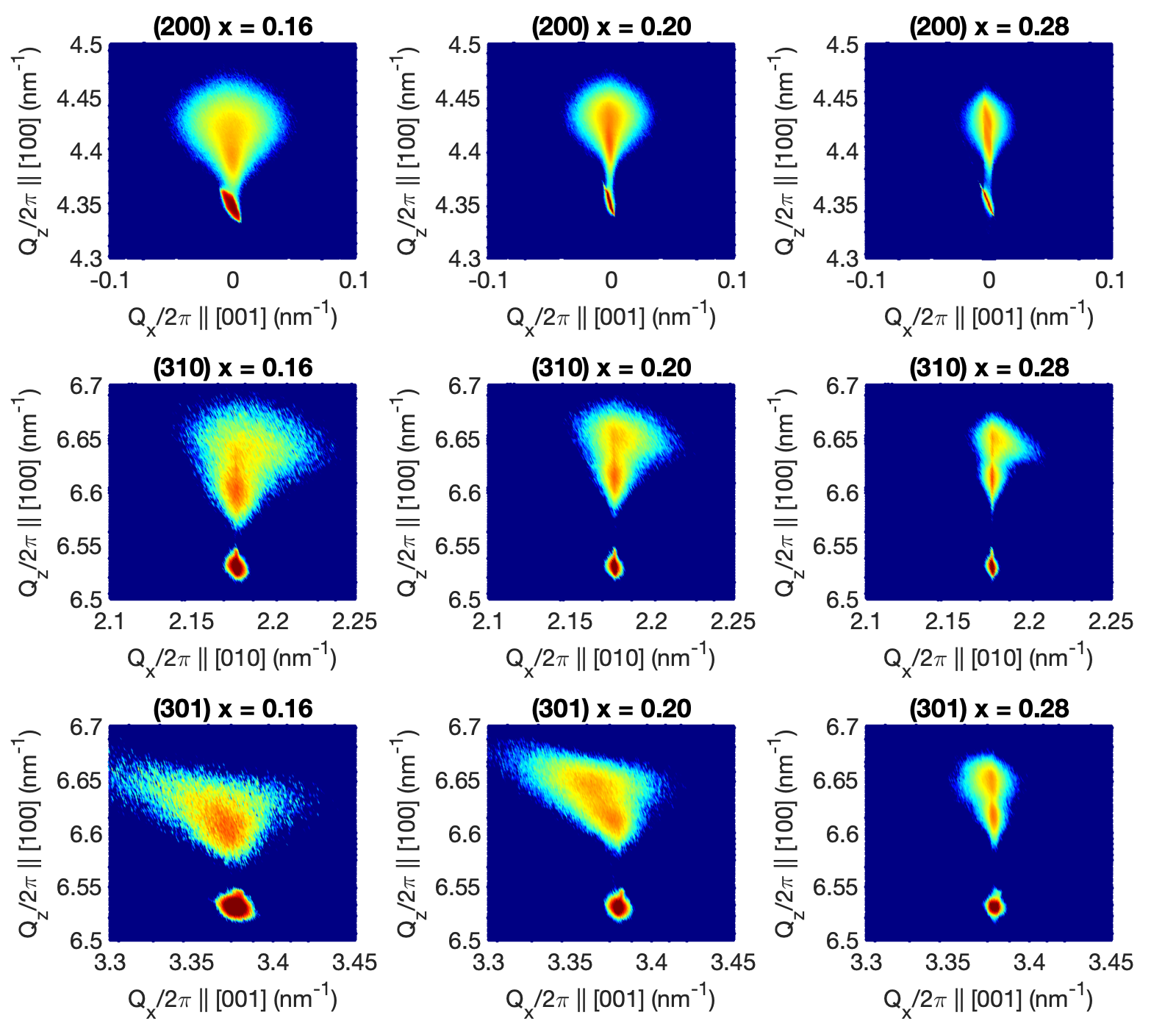}
\caption{\textbf{RSM's of (100) oriented samples:} (a)-(c) (200) peaks of $x=0.16,0.20,0.28$, (d)-(f) (310) peaks of $x=0.16,0.20,0.28$, and (g)-(i) (301) peaks of $x=0.16,0.20,0.28$.}
\label{suppfig:2}
\end{figure*}

\begin{figure*}[h]
\centering \includegraphics[width=1\columnwidth]{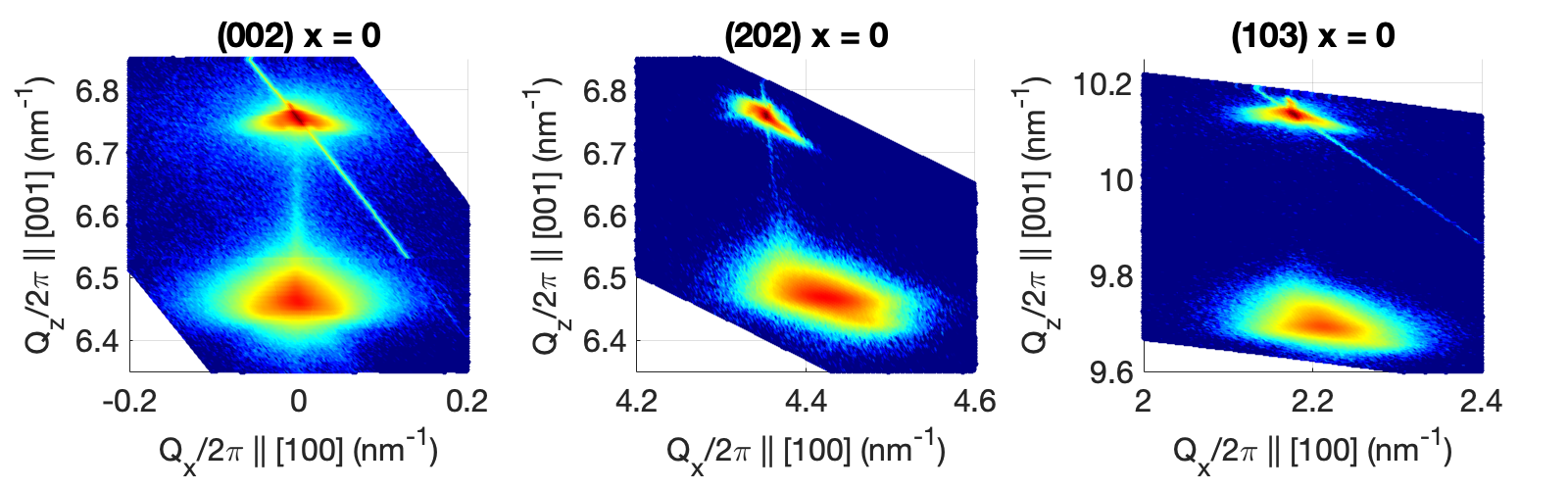}
\caption{\textbf{RSM's of (001) oriented sample with $x=0$:} (a) (002) peak, (b) (202) peak, and (c) (103) peak.}
\label{suppfig:3}
\end{figure*}

\begin{figure*}[h]
\centering \includegraphics[width=1\columnwidth]{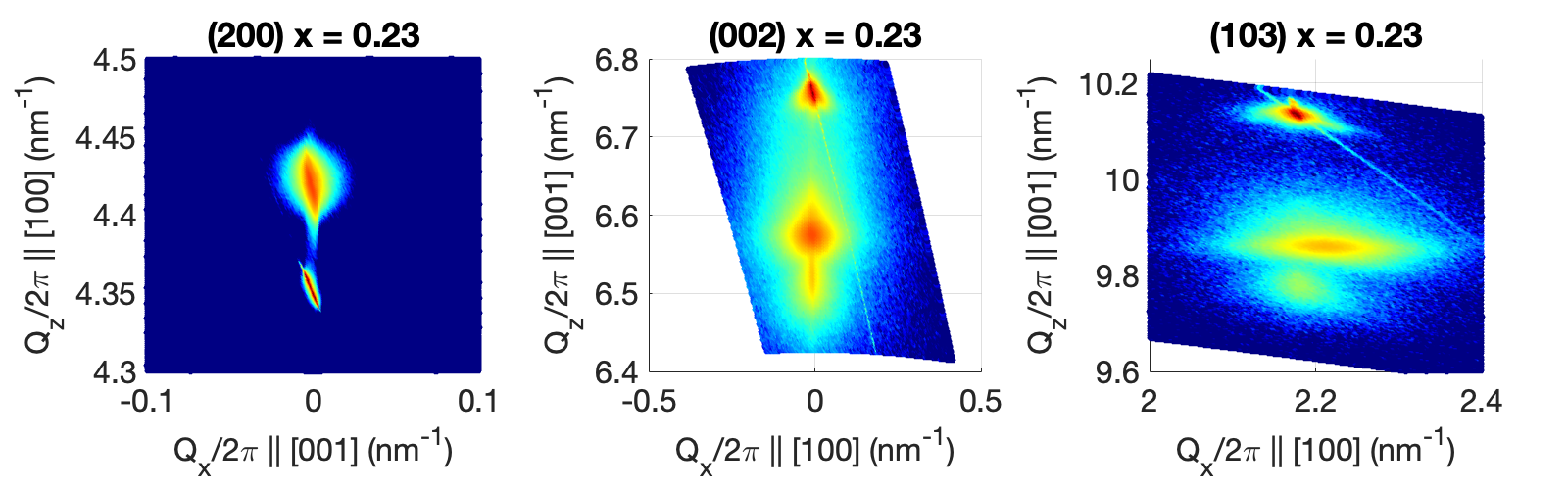}
\caption{\textbf{Additional RSM's of $x=0.23$ samples from main text:} (a) (200) peak of (100) sample, (b) (002) peak of (001) sample, and (c) (103) peak of (001) sample.}
\label{suppfig:4}
\end{figure*}

RSMs of the (100) oriented samples with $x = 0.16, 0.20,$ and $0.28$ are shown in Figure \ref{suppfig:2}, and RSMs of the (001) oriented $x = 0$ sample are shown in Figure \ref{suppfig:3}. For the (100) and (001) sample with $x = 0.23$, additional RSMs not included in the main text are shown in Figure \ref{suppfig:4}. A summary of the material properties of all samples, including average composition, lattice parameters, phase fraction, and thickness, are reported in Table \ref{tab:two_column_table}. The phase fraction reported is from the intensity ratio of the different diffraction peaks for a strained and relaxed phase, and it was used to convert from volume susceptibility to molar susceptibility.

\begin{table*}[h] 
\centering
\caption{Sample Properties}
\label{tab:two_column_table}

\begin{tabular}{|c|c|c|c|c|c|c|c|}
\hline
\textbf{(hkl)} & \textbf{Sample} & \textbf{Cr $x$} & \textbf{$a$ (\AA)} & \textbf{$b$ (\AA)} & \textbf{$c$ (\AA)} & \textbf{Phase fraction} & \textbf{thickness (nm)} \\ \hline
\multirow{2}{*}{(001)} 
& RuO\textsubscript{2} & $0 \pm 0$ & $4.520$ & $4.520$ & $3.091$ & 1 & $200$ \\ \cline{2-8} 
& RuO\textsubscript{2} & $0 \pm 0$ & $4.586$ & $4.586$ & $3.063$ & 1 & $19.5$ \\ 
& Ru\textsubscript{0.77}Cr\textsubscript{0.23}O\textsubscript{2} & $0.231 \pm 0.002$ & $4.517$ & $4.517$ & $3.039$ & 1 & $160$ \\ \hline
\multirow{6}{*}{(100)} 
& Ru\textsubscript{0.84}Cr\textsubscript{0.16}O\textsubscript{2} 
& $0.156 \pm 0.002$ & $4.546$ & $4.592$ & $2.964$ & 1 & $210$ \\ \cline{2-8} 

& \multirow{2}{*}{Ru\textsubscript{0.80}Cr\textsubscript{0.20}O\textsubscript{2}} 
& \multirow{2}{*}{$0.196 \pm 0.003$} 
& 4.537 & 4.593 & 2.962 & 0.53 & \multirow{2}{*}{$177$} \\ 

& & & 4.515 & 4.589 & 2.973 & 0.47 & \\ \cline{2-8}

& Ru\textsubscript{0.77}Cr\textsubscript{0.23}O\textsubscript{2} 
& $0.227 \pm 0.003$ & $4.527$ & $4.592$ & $2.958$ & 1 & $141$ \\ \cline{2-8}

& \multirow{2}{*}{Ru\textsubscript{0.72}Cr\textsubscript{0.28}O\textsubscript{2}} 
& \multirow{2}{*}{$0.279 \pm 0.004$} 
& 4.536 & 4.591 & 2.960 & 0.54 & \multirow{2}{*}{$154$} \\ 

& & & 4.511 & 4.591 & 2.961 & 0.46 & \\ \hline
\end{tabular}
\end{table*} 

\begin{figure*}
\centering \includegraphics[width=0.7\columnwidth]{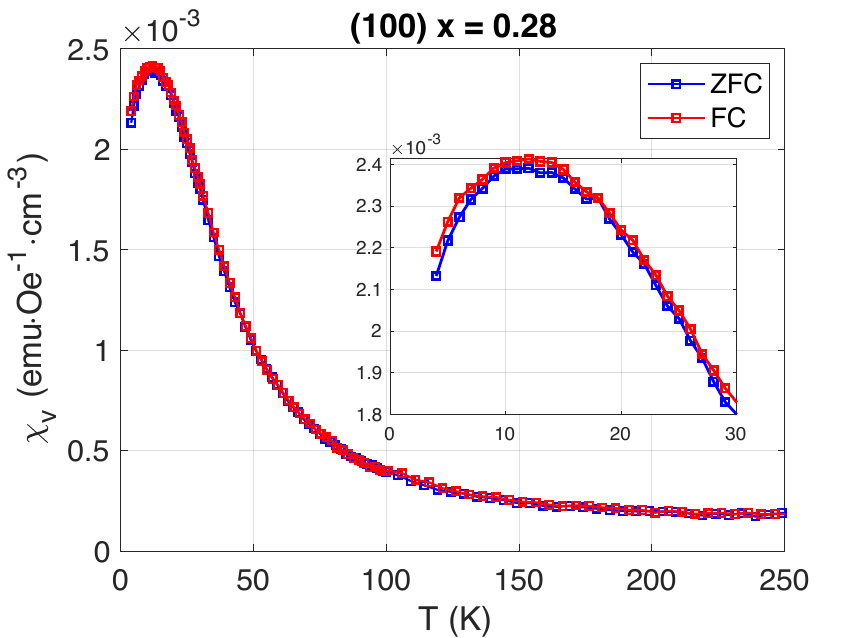}
\caption{\textbf{Comparison of zero-field cooled and field cooled susceptibility:} volume susceptibility of the (100) $x = 0.28$ sample, with external field of $1~$T oriented along the [001] direction.}
\label{suppfig:5}
\end{figure*}

The susceptibility of all Cr-alloyed samples was measured using both a zero-field cool (ZFC) and field cool (FC) procedure. As there is negligible difference between the two measurements, we report the average of the ZFC and FC curves in the main text to improve the signal-to-noise ratio. An example of the ZFC and FC curves is given for the (100) sample with $x = 0.28$ in Figure \ref{suppfig:5}.

\providecommand{\noopsort}[1]{}\providecommand{\singleletter}[1]{#1}%

\end{document}